# Quantifying the radiative response to surface temperature variability: A critical comparison of current methods

Leif Fredericks,[a] Maria Rugenstein,[a] David W.J. Thompson,[a,b] Senne Van Loon,[a] Fabrizio Falasca,[c] Rory Basinski-Ferris,[d] Paulo Ceppi,[e] Quran Wu,[f] Jonah Bloch-Johnson,[f,g] Marc Alessi,[a,h] and Sarah M. Kang[i]

[a] *Department of Atmospheric Science, Colorado State University, Fort Collins, CO, USA*

[b] *School of Environment Sciences, University of East Anglia, Norwich, UK*

[c] *Courant Institute School of Mathematics, Computing and Data Science, New York University, New York, NY, USA*

[d] *Scripps Institution of Oceanography, University of California San Diego, La Jolla, CA, USA*

[e] *Department of Physics, Imperial College London, London, UK*

[f] *National Centre for Atmospheric Science, University of Reading, Reading, UK*

[g] *Department of Earth and Climate Sciences, Tufts University, Medford, MA, USA*

[h] *Climate & Energy Program, Union of Concerned Scientists, Washington, D.C., USA*

[i] *Max Planck Institute for Meteorology, Hamburg, Germany*

*Corresponding author*: Leif Fredericks, leif.fredericks@gmail.com

ABSTRACT: Over the past decade, it has become clear that the radiative response to surface temperature change depends on the spatially varying structure in the temperature field, a phenomenon known as the "pattern effect". The pattern effect is commonly estimated from dedicated climate model simulations forced with local surface temperatures patches (Green's function experiments). Green's function experiments capture causal influences from temperature perturbations, but are computationally expensive to run. Recently, however, several methods have been proposed that estimate the pattern effect through statistical means. These methods can accurately predict the radiative response to temperature variations in climate model simulations. The goal of this paper is to compare methods used to quantify the pattern effect. We apply each method to the same prediction task and discuss its advantages and disadvantages. Most methods indicate large negative feedbacks over the western Pacific. Over other regions, the methods frequently disagree on feedback sign and spatial homogeneity. While all methods yield similar predictions of the global radiative response to surface temperature variations driven by internal variability, they produce very different predictions from the patterns of surface temperature change in simulations forced with increasing $CO_2$ concentrations. We discuss reasons for the discrepancies between methods and recommend paths towards using them in the future to enhance physical understanding of the pattern effect.

## 1. The radiative response to surface temperature change

The global energy balance of the Earth can be expressed as:

$$N = F + R, \tag{1}$$

where $F$ represents the radiative forcing on the system relative to equilibrium, including anthropogenic forcings due to increases in $CO_2$ and aerosols, or natural forcings due to volcanic eruptions; $N$ represents the net radiative imbalance between incoming and outgoing radiation at the top of the atmosphere, which is mostly stored in the ocean; and $R$ represents the radiative response of the Earth system to the forcing. In an equilibrium state, radiative energy is balanced so that $N = 0$. However, when the system is taken out of equilibrium due to a forcing ($F$), the radiative imbalance ($N$) will be nonzero until balanced by the Earth's radiative response ($R$) at a new equilibrium.

All three terms are important for understanding climate change, and are fields of study in their own right. The magnitude of the forcing ($F$) determines how strongly the Earth's energy balance is displaced from equilibrium (e.g., Arrhenius 1896; Ramanathan et al. 1979; Hansen et al. 1981; National Research Council 2005; Raghuraman et al. 2021; Ramaswamy et al. 2019). The uptake of heat by the oceans, which accounts for approximately 90% of the energy imbalance ($N$), dictates how effectively the Earth buffers the radiative imbalances caused by such forcing and sets the time to reach a new equilibrium (e.g., Gregory 2000; Hansen et al. 2005, 2011; Trenberth et al. 2014; Zanna et al. 2019). The radiative response ($R$) determines how much energy is emitted back to space in response to the forcing. Climate sensitivity quantifies the equilibrium surface temperature in response to a specified $CO_2$ forcing and depends heavily on $R$.

Here we focus on one key aspect of the radiative response $R$: how variations in the spatial pattern of surface temperature affect global-mean radiation. This dependency is called the "pattern effect" on radiation.

## 2. Challenges in estimating $R$

The radiative response is determined by the amplitude of radiative feedbacks operating within the climate system. It has long been approximated (e.g., Hansen et al. 1997; Senior and Mitchell 2000; Gregory et al. 2004; Sherwood et al. 2020) as the product of a global feedback parameter ($\lambda$) and the global surface temperature change ($\Delta T_s$). In this case, $R = \lambda \Delta T_s$ and the energy balance in Eq. 1 is

$$N = F + \lambda \Delta T_s. \tag{2}$$

In this linear approximation, equilibrium climate sensitivity (ECS)—an estimate for the new equilibrium surface temperature following a forced departure from the previous equilibrium—can be estimated for an instantaneous doubling of carbon dioxide ($F_{2xCO_2}$) by setting $N = 0$ and solving for the change in temperature:

$$\mathrm{ECS} = -\frac{F_{2xCO_2}}{\lambda}. \tag{3}$$

Equation 3 provides a convenient climate change assessment derived from the global feedback parameter, $\lambda$. In the last decade, however, it has become increasingly clear that the linear formulation $R = \lambda \Delta T_s$ is insufficient to describe the global radiative response. Numerous studies indicate a time-dependency in $\lambda$—and thus the climate sensitivity—when coupled climate models are run towards equilibrium following a constant forcing $F$ (e.g., Senior and Mitchell 2000; Winton et al. 2010; Andrews et al. 2015; Proistosescu and Huybers 2017; Paynter et al. 2018; Rugenstein et al. 2019, 2020). Similarly, atmospheric models forced with historical surface temperatures reveal substantial variations in $\lambda$ over time (e.g., Zhou et al. 2016; Andrews and Webb 2018; Andrews et al. 2022). There is also evidence that the spatial pattern of forcing can affect $\lambda$ (e.g., Zhang et al. 2023a), though we do not address this complicating factor in the current work.

The dependency of $R$ on the specific pattern of surface temperature anomalies is referred to as the "pattern effect" (e.g., Andrews et al. 2015; Stevens et al. 2016). The implication is that $R$ is a function of many regional temperature anomalies rather than a single global surface temperature anomaly as in Eq. 2.

The dependence of $R$ on the spatial pattern of temperature change can be approximated from the chain rule as follows:

$$R \approx \frac{\partial R}{\partial T_1}\Delta T_1 + \frac{\partial R}{\partial T_2}\Delta T_2 + \ldots \frac{\partial R}{\partial T_i}\Delta T_i = \beta_1 \Delta T_1 + \beta_2 \Delta T_2 + \ldots \beta_i \Delta T_i, \tag{4}$$

where $\partial R/\partial T_i$ represents the partial derivative of global radiation to a temperature change at location $i$, $\Delta T_i$ represents a temperature perturbation at $i$, and $\beta_i$ reformulates the partial derivatives as regression coefficients, as estimated by a particular method.

Radiation also has spatial variation, and $R$ (global mean radiation) is the mean across the spatial radiative field. This highlights the main complication in representing $R$. The radiative response at a given location $j$ ($R_j$) can depend on both a) the local influence of temperature at the same location and b) the remote influence of temperatures elsewhere on the planet.

For example, the local radiative properties in the southeastern tropical Pacific (SEP) are thought to be set by processes driven by both local and remote surface temperature. This region is significant for the climate because the broad region of marine stratocumulus clouds in the SEP has a notable impact on global radiation (e.g., Bony and Dufresne 2005; Zelinka et al. 2020; Myers et al. 2021, 2023; Ceppi et al. 2024; Breul et al. 2025). The cloud fraction and thus net shortwave cloud radiative effect in this region is a function of both the local surface temperature and inversion strength. The local effect of warming in the SEP is to decrease the local inversion strength and thus lower cloud fraction. However, remote temperature variations also affect inversion strength in this region.

One remote influence comes from the west Pacific, which appears to be a dominant region of negative feedback. A warming in the west Pacific has a cooling effect from a reduction in global $R$ (e.g., Dong et al. 2019). Warming in regions of deep convection in the west Pacific is associated with increases in inversion strength and higher cloud fraction in the SEP (Zhou et al. 2016, 2017; Dong et al. 2019; Andrews and Webb 2018; Mauritsen 2016). In fact, the cloud radiative effect in the SEP responds more efficiently to west Pacific warming than to local SEP warming (Alessi and Rugenstein 2023). The SST gradient between the east and west Pacific can also act through convective aggregation to expand or contract large-scale dry and clear-sky regions. This modulates the strength of global outgoing longwave radiation (e.g., Quan et al. 2025). Another remote influence comes from the Southern Ocean, which exerts a positive feedback on the SEP (Kang et al. 2023). Cooling in the extratropics is linked to surface cooling in the SEP, which strengthens the inversion and increases low marine clouds (e.g., Kang et al. 2020; Kim et al. 2022; Zheng et al. 2025).

Hence local feedbacks on global radiation provide an incomplete accounting of the relationships between temperature and radiation (e.g., Bloch-Johnson et al. 2020; Hedemann et al. 2022). This mirrors the argument against using global temperature to predict global radiation, and indeed Eq. 4 can be reformulated for radiation at a specific location as:

$$R_j \approx \frac{\partial R_j}{\partial T_1}\Delta T_1 + \frac{\partial R_j}{\partial T_2}\Delta T_2 + \ldots \frac{\partial R_j}{\partial T_i}\Delta T_i = \beta_{1,j}\Delta T_1 + \beta_{2,j}\Delta T_2 + \ldots \beta_{i,j}\Delta T_i. \quad (5)$$

Here, radiation at each location $j$ depends on temperature at each location $i$. In this work, we predict global radiation directly, which is equivalent to predicting the coefficients in Eq. 4. However, we emphasize that the effect of any given $T_i$ on global radiation ($R$) is the sum of how each $T_i$ affects each $R_j$. It is possible to use any of the methods to estimate a sensitivity map for every $R_j$, but that is beyond the scope of this comparison.

## 3. Causal methods to estimate $R$

The contribution of the spatial structure of temperature change to $R$ can be quantified in numerical experiments using the so-called "Green's function approach". In this case, the partial derivatives in Eq. 4 are explicitly estimated. A Green's function experiment perturbs temperature over large patches of the surface, typically elliptical in shape, and produces a map indicating the weighted average of these responses at each location. To isolate the effects of a single patch, the perturbed experiments are compared to a control simulation, which records the response in a climate model to a prescribed time series of unperturbed sea surface temperatures. The effects of temperature change in that patch are thus the difference between the equilibrated response in the patch experiment and the control.

Patch simulations are repeated for patches covering the ocean grid across the full globe. This directly quantifies the causal sensitivity of global mean radiation to temperature perturbations at each location independently (Zhou et al. 2017, 2020; Dong et al. 2019; Zhang et al. 2023b; Alessi and Rugenstein 2023; Quan et al. 2024; Bloch-Johnson et al. 2024). A "sensitivity map" is a spatial plot of these values, indicating the partial contribution of local temperature to global radiation. In this way it is a spatial plot of the $\partial R/\partial T_i$ terms in Eq. 4. Projecting a sensitivity map onto a map of temperature anomalies ($\Delta T_i$) estimates a value for global radiation consistent with the patch responses.

While providing a causal solution to Eq. 4, the Green's function is computationally expensive, requiring thousands of simulated years. The results are also dependent on the specific protocol employed. For example, differences in patch size, distribution, and overlap can lead to different estimated feedbacks (discussed later). In terms of representing $R$, the Green's function is also restricted to describing whichever climate model it has been estimated from, and cannot be applied directly to observations. Any biases present in the climate model's radiative response will propagate to the Green's function.

To properly capture a representation of $R$ in the real world, we need to incorporate observations. However, methodological constraints like those of the Green's function are not the only challenges to doing so. Satellites observe the difference between incoming and outgoing radiation (i.e. $N$, Eq. 1), so $R$ is not directly observable. Reliable observations of $N$ only extend back to 2001, so the length of the observational record is short relative to known timescales of multidecadal variations. This time period has also been influenced by a combination of internal variability [e.g., the El Niño-Southern Oscillation (ENSO) and stochastic weather] and an uncertain degree of anthropogenic forcing (e.g., Forster et al. 2021; Van Loon et al. 2025a). There is also no way to distinguish whether variations in observed $N$ at a particular location are due to local temperature variations, remote influences from temperature elsewhere, or even spatial inhomogeneity in the forcing, $F$.

As such, in the last few years, an increasing number of methods have been proposed to approximate the linear system in Eq. 4 that leverage statistical relationships within existing simulations of internal variability. These methods are computationally cheaper than Green's function experiments. They can also in principle be applied to observations (e.g., Thompson et al. 2025; Fredericks et al. 2025), although most have focused on testing on models to date.

## 4. Comparing statistical estimates of $R$

In this section we compare 1) estimates of the global mean radiative response provided by applying different statistical methods to the relationships between the global-mean radiative flux $R$ and the spatially-varying temperature field $T_i$ with 2) estimates of the global mean radiative response as given by Green's function experiments. These methods predict global radiation either with linear sensitivity maps, as discussed for the Green's function, or with data-trained nonlinear operations. The different methods and guiding equations are summarized in Fig. 1 and include:

- Maximum covariance analysis (MCA; Thompson et al. 2025): MCA defines the sensitivity map which, when projected back onto the training pattern in temperature, predicts the radiative time series that covaries maximally with the actual $R$.
- Ordinary least squares (OLS; Bloch-Johnson et al. 2020): Ordinary least squares generates a sensitivity map from regression coefficients which, when projected back onto the training pattern in temperature, minimizes squared error between the actual and predicted $R$.
- Regularized regression (ridge, LASSO, and elastic net; Fredericks et al. 2025): Regularized regression adds additional constraint(s) to the general OLS optimization to reduce overfitting. Ridge regression penalizes the sum of squared magnitudes in the vector containing the regression coefficients. LASSO instead penalizes the sum of absolute magnitudes. Elastic net incorporates both penalties.
- Principal component (PC) regression (Wu 2026): Principal component regression generates a sensitivity map by regressing global radiation onto time series of empirical orthogonal functions (EOFs) rather than onto grid cell temperatures to reduce dimensionality.
- Partial least squares (PLS): PLS analysis iteratively applies a covariance analysis analogous to MCA. The PLS sensitivity map is a linear combination of the sensitivity maps across all iterations.
- Fluctuation-dissipation relation (FDR; Falasca et al. 2025): FDR generates a spatiotemporal response operator from the training data. It can be applied both to behave in an atmosphere-only configuration that approximates the sensitivity map of a traditional Green's function, or in a coupled-dynamics configuration that allows the surface to respond as well as the atmosphere.
- Convolutional neural network (CNN; Rugenstein et al. 2025; Van Loon et al. 2025a): A CNN uses training data to fix weights in 2-dimensional kernels. When convolved with maps of surface temperature anomalies, the kernels create secondary, feature extraction maps. The process is performed iteratively, with nonlinear functions applied to the maps between convolutional steps.

- Green's function (Zhou et al. 2017, 2020; Dong et al. 2019; Alessi and Rugenstein 2023; Zhang et al. 2023b; Bloch-Johnson et al. 2024): The Green's function generates a sensitivity map from a series of climate model patch perturbation experiments covering the global ocean. Sensitivities are calculated from the difference between simulations with a perturbation at a given location and a control simulation.

- AMIP ridge regression (Kang et al. 2023; Ceppi and Nowack 2021): This application of ridge regression first uses AMIP simulations to estimate the global feedback, then generates a sensitivity map for the pattern component of the radiative response from regression coefficients. The global feedback predicts the global mean change in $R$, and the sensitivity map predicts the additional contribution of the temperature pattern to $R$.

We have trained all methods on ~1,000 years of annual mean spatial surface temperature and global net radiation in a pre-industrial control (piControl) simulation. This is a climate model experiment where atmospheric conditions are set to 1850 conditions so the climate system experiences internal variability without anthropogenic forcing. Although it may limit methods that could take advantage of feedbacks on shorter timescales, we use annual rather than monthly means because they more closely approximate an equilibrium response in $R$ to sustained anomalies in $T_i$. We use the Max Planck Institute Earth System Model v1.2 (MPI-ESM1.2; Rohrschneider et al. 2019; Mauritsen et al. 2019). The Green's function and AMIP ridge regression are an exception to the training process. The Green's function instead uses many patch perturbation simulations as discussed, and uses the same atmospheric model as MPI-ESM1.2. AMIP ridge regression uses three simulation protocols: the standard AMIP simulation (Gates et al. 1999); a protocol otherwise identical to AMIP but with a uniform 4-Kelvin increase in sea surface temperature (SST), AMIP-p4k; and an AMIP-style simulation from 1870 to 2014 that maintains the atmospheric conditions of the piControl throughout the simulation, AMIP-piForcing. These also use the atmospheric model employed in MPI-ESM1.2.

In Fig. 2, we show how well each of the methods predicts internal variability. Our test is a 100-year period taken from the piControl simulation that is not included in the ~1,000 years of training data. This test is "out-of-sample", but "in-distribution" because it includes the same conditions as the training data. The test consists of applying each method to the spatial temperature time series and predicting the global radiative response, $R(t)$. By comparing the method-predicted radiative

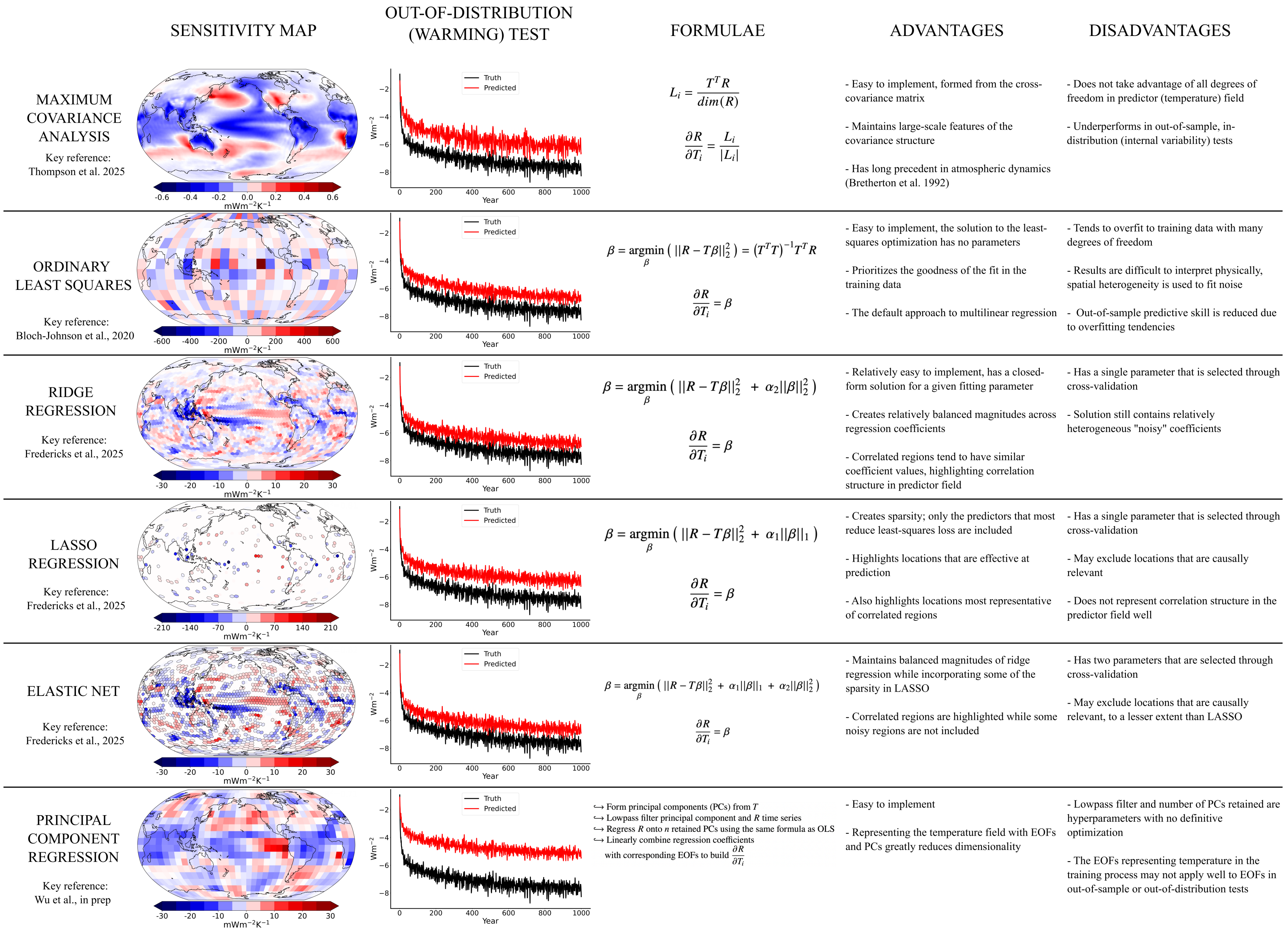

FIG. 1. Comparison of methods, part 1.

| | SENSITIVITY MAP | OUT-OF-DISTRIBUTION (WARMING) TEST | FORMULAE | ADVANTAGES | DISADVANTAGES |
|---|---|---|---|---|---|
| PARTIAL LEAST SQUARES | 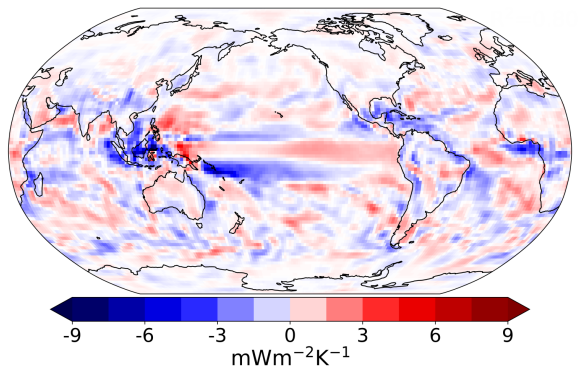 | 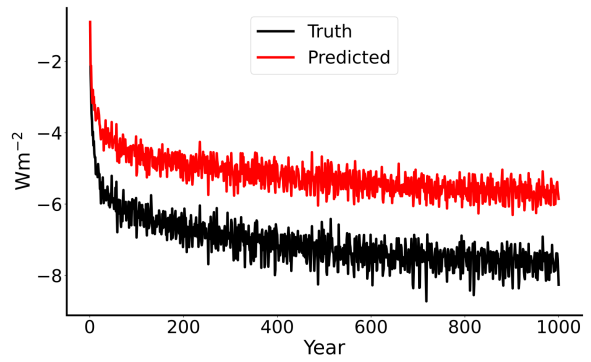 | ↪ Compute the left singular matrix $u_k$ of $T^T R$ (i.e., MCA)<br>↪ Regress out the influence of $u_k$ from $T$ and $R$<br>↪ Iteratively repeat $n$ times<br>↪ $\frac{\partial R}{\partial T_i}$ is the linear combination of $n$ components | - Comprised of components with physical meaning, i.e., essentially iterative MCA<br>- Similar to ridge, reduces overfitting tendencies with balanced coefficients<br>- Integer parameter (number of retained components) is much easier to optimize than a rational parameter | - Has a single parameter that is selected through cross-validation<br>- Must be implemented via iterative algorithm<br>- Different variations of PLS exist due to different implementations of the algorithm |
| FLUCTUATION-DISSIPATION RELATION: COUPLED<br>Key reference:<br>Falasca et al., 2025 | 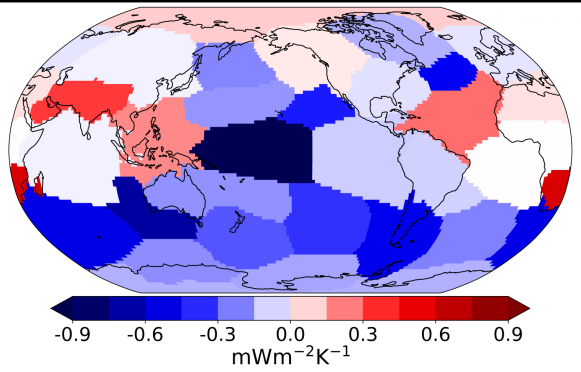 | 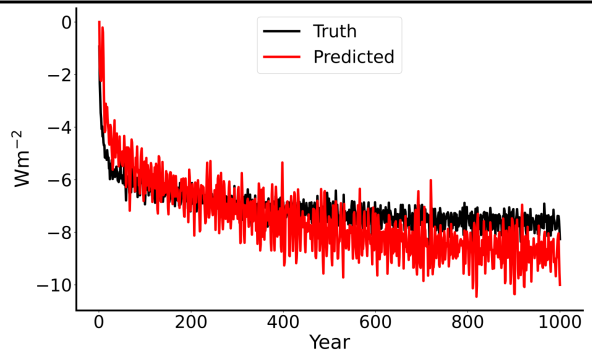 | ↪ Reduce $T$ into regionally constrained patterns<br>↪ Combine $T$ and $R$ into a single state vector $\boldsymbol{x}(\boldsymbol{t})$<br>↪ Advance $\boldsymbol{x}(\boldsymbol{t})$ via the convolution:<br>$\delta\langle \boldsymbol{x}(\boldsymbol{t})\rangle = \int_0^t \mathcal{R}(\tau)\delta x(t-\tau)d\tau$<br>↪ $\mathcal{R} = \boldsymbol{C}(t)\boldsymbol{C}(0)^{-1}$; $\boldsymbol{C}_{i,j} = \langle \boldsymbol{x}_i(\tau+t)\boldsymbol{x}_j(\tau)\rangle$<br>↪ $\delta x(t)$ are prescribed surface temperatures applied as perturbations<br>↪ $\tau$ is the response time scale<br>↪ Map shows **cumulative** response to step function perturbations applied independently to regions | - Estimates a Green's function-like perturbation experiment in coupled, rather than atmospheric-only, climate models<br>- Captures feedback effects across multiple timescales by including lagged responses in the convolution, accounting for atmospheric-ocean interactions such as teleconnection patterns | - Relatively complex interpretation, the sensitivity map is not a linear description of the prediction process<br>- Depends on careful considerations in the coarse-graining procedure, including the focus on large spatial averages rather than grid-scale processes |
| FLUCTUATION-DISSIPATION RELATION: ATMOSPHERIC<br>Key reference:<br>Falasca et al., 2025 | 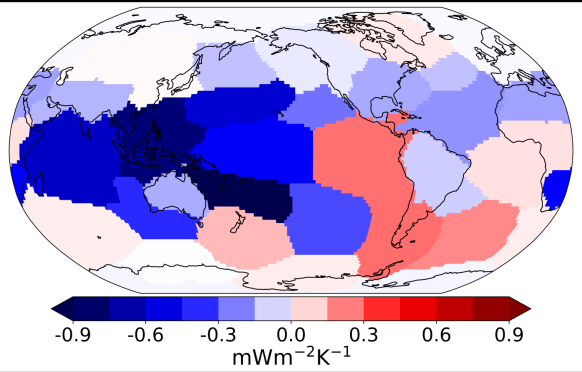 | 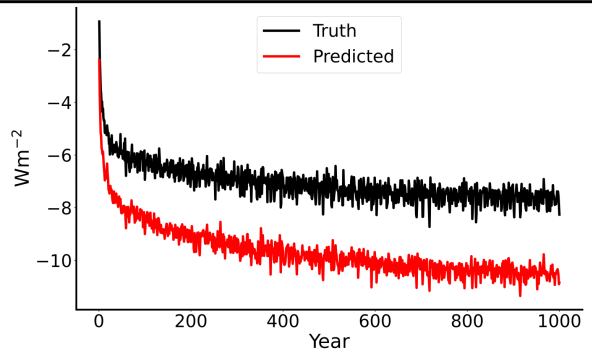 | ↪ Define $\boldsymbol{x}(\boldsymbol{t})$ and $\mathcal{R}$ with $\tau$ at 1 month<br>↪ Apply a step temperature anomaly to a single region<br>↪ Calculate the response in $R$ at 1 month<br>↪ Repeat independently for each pattern, as in coupled<br>↪ $\frac{\partial R}{\partial T_i}$ is then approximated by using $\mathcal{R}$ truncated at 1 month to capture the fastest timescales | - Approximates a Green's function experiment with much less computational cost<br>- Sensitivity map has a standard interpretation because applied temperature anomalies are effectively boundary conditions<br>- Tailored to the problem of prediction and reconstruction | - Limited to estimate responses to temperature perturbations in an atmospheric-only system<br>- Similar to Green's function experiments it underperforms in out-of-sample, in-distribution (internal variability) tests compared to FDR-C<br>- Similarly to FDR-C, depends on careful considerations in the coarse-graining procedure |
| CONVOLUTIONAL NEURAL NETWORK<br>Key references:<br>Rugenstein et al., 2025;<br>Van Loon et al., 2025 | 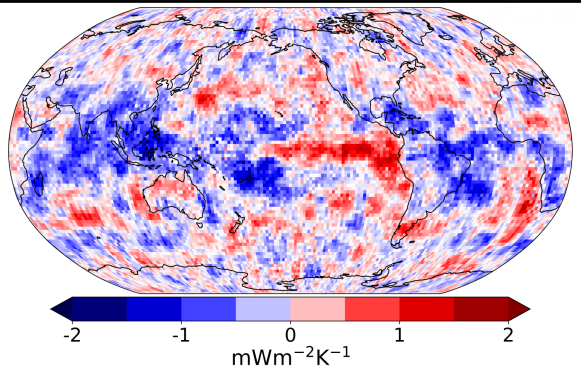 | 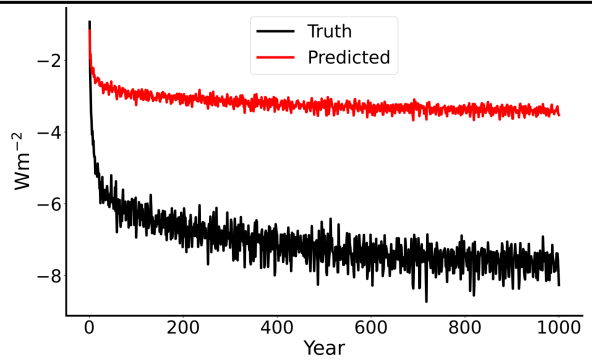 | ↪ Define kernels to convolve over 2-D $T$ pattern<br>↪ Repeat through multiple layers<br>↪ Reduce to a single value via 1-D neural net<br>↪ Train weights in kernels and conections on $T$ and $R$<br>↪ To build map, identify gradient from each $T_i$ to $R$<br>↪ Map the average gradient across predicted patterns | - Accounts for nonlinearities in the temperature fields, making the CNN more adaptive within distribution<br>- Can incorporate improvements from the rapidly evolving field of machine learning<br>- Applications to related problems can improve skill through transfer learning | - Requires selecting several hyperparameters; dependent on large training data sets<br>- Difficult to visualize, the sensitivity map is an average of the assessed sensitivities across all predicted temperature patterns<br>- Performance in out-of-distribution test cases requires more training data than included here |
| GREEN'S FUNCTION<br>Key references:<br>Zhou et al., 2017;<br>Dong et al., 2019;<br>Zhou et al., 2020;<br>Alessi and Rugenstein 2023<br>Zhang et al., 2023;<br>Bloch-Johnson et al., 2024 | 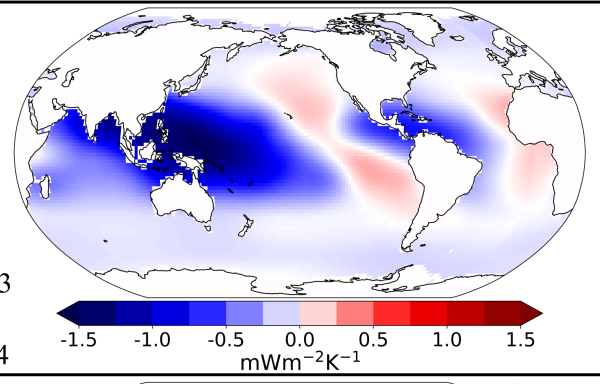 | 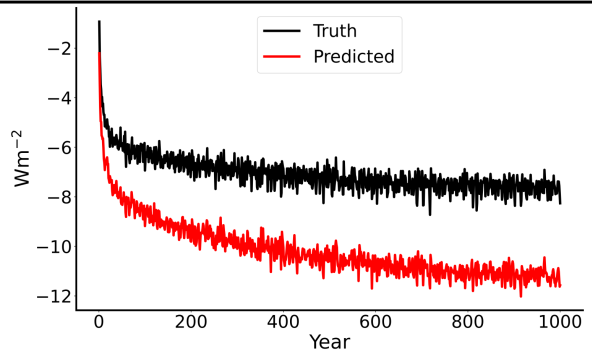 | ↪ Control run: atmospheric GCM with prescribed SSTs<br>↪ Add constant patch perturbation to control run<br>↪ Determine equilibrated response in $R$ to patch run<br>↪ Compare to control to find partial contribution<br>↪ Repeat for patches covering the global ocean<br>↪ Calculate linear influence from each grid point to determine $\frac{\partial R}{\partial T_i}$ | - By using the radiative response to independent patches perturbations, represents true partial derivative from radiation to local temperature<br>- Only causal method considered<br>- Mechanisms can be investigated from individual patch simulations | - Computationally expensive to simulate each patch perturbation<br>- Captures responses to unrealistic temperature patterns, which have been shown to act nonlinearly in combination<br>- Must be tuned to capture the forced response in the warming test |
| AMIP RIDGE REGRESSION<br>Key references:<br>Kang et al., 2023;<br>Ceppi and Nowack, 2021 | 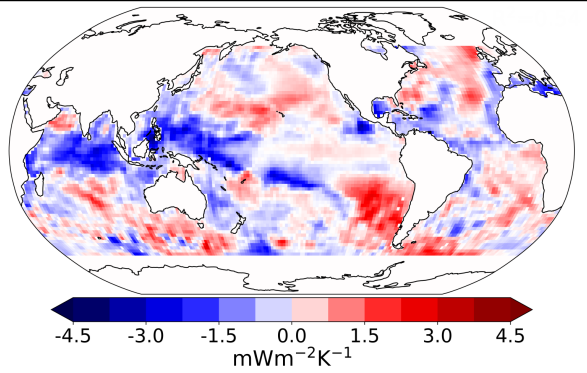 | 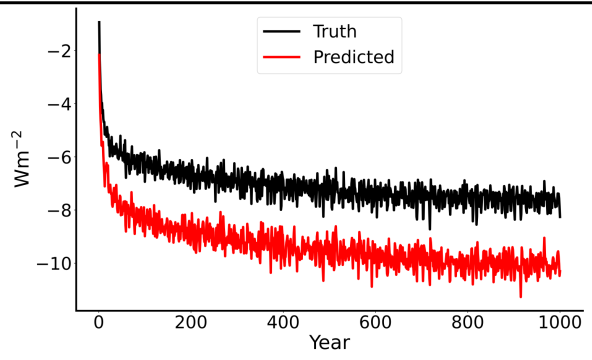 | ↪ $SST_{pattern}(r,t) = SST(r,t) - SST_{mean}(t)$ in amip-piForcing<br>↪ $dR_{4K}$ = difference in $R$ between amip and amip-p4k<br>↪ Calculate $R_{mean}(t) = \frac{dR_{4K}}{4} \times SST_{mean}(t)$ in amip-piForcing<br>↪ Pattern response: $R_{pattern}(t) = R(t) - R_{mean}(t)$<br>↪ Ridge regress $R_{pattern}(t)$ onto $SST_{pattern}(r,t)$<br>↪ Regression coefficients $\frac{\partial R}{\partial T_i}$ predict the pattern component, $dR_{4K}$ predicts the global component. | - Using amip ensures the temperature patterns in the training data actually occur in the real world<br>- Incorporates forced response in the training with relatively less computational cost than the Green's function | - Cannot be assessed from existing piControl simulations. Requires: amip-piForcing, amip, and amip-p4k simulations<br>- Unlike the Green's function, not a true causal representation of partial derivatives. Coefficients are statistical connections as in the multilinear regression methods |

Fig. 1. Comparison of methods, part 2 (caption on following page).

Fig. 1. (Previous page.) Comparison of methods: For each method ordered approximately from simplest to most complex process: 1) The estimated sensitivity from the global radiative response to temperature at each location, given in units of milliWatts per square meter per Kelvin (magnitudes are not comparable from one map to another due to differing resolution and masking); 2) the radiative response predicted by each method in a 4x$CO_2$ warming experiment plotted against the simulated truth; 3) a summary of the formulae or procedures used; 4) advantages of the method; and 5) disadvantages of the method.

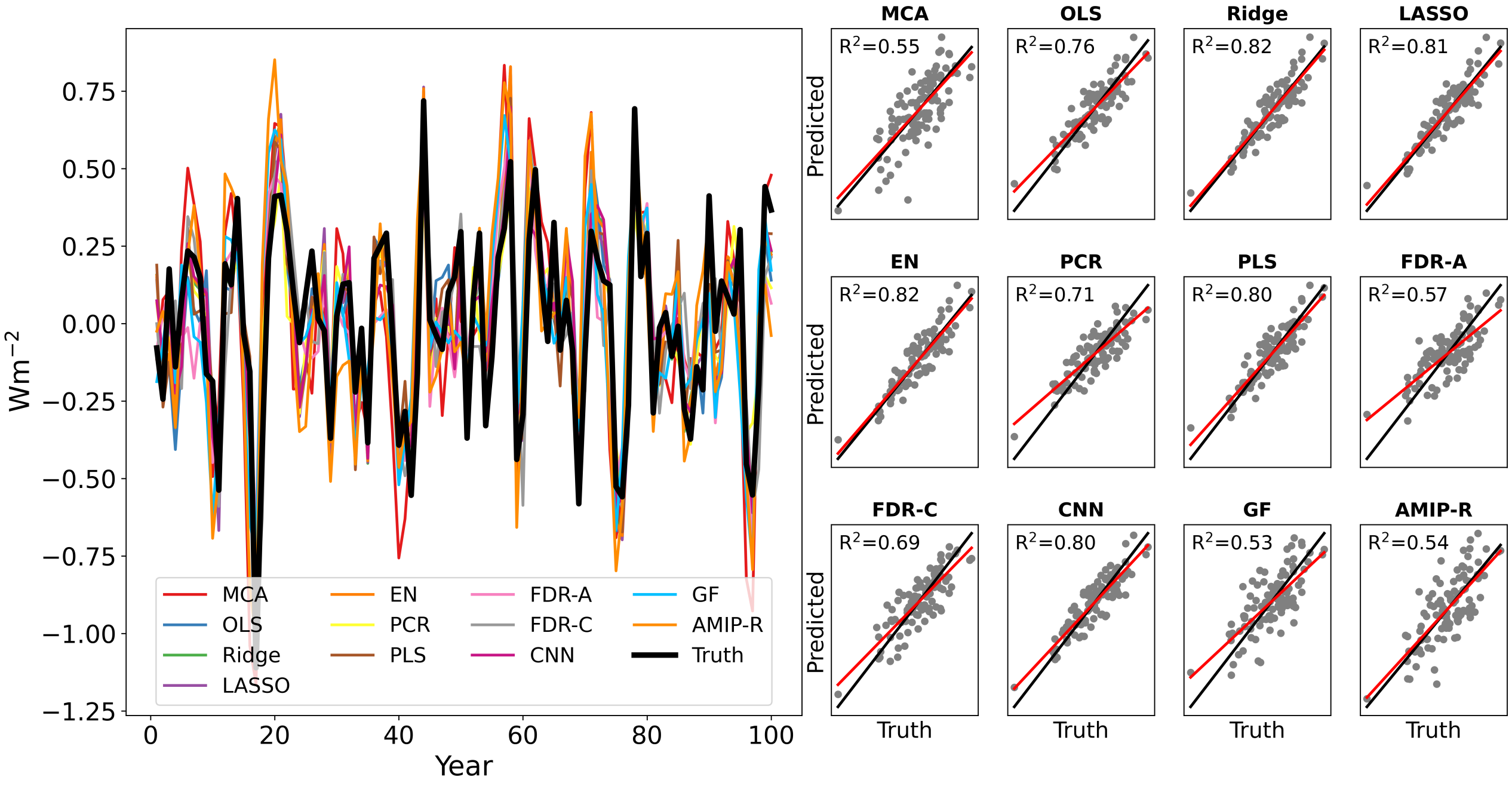

Fig. 2. Performance of each method at predicting $\hat{R}(t)$ in 100 years of piControl internal variability withheld from the training data, shown as time series of predictions vs. simulated truth (left) and as regressions of predicted values onto simulated values (right). Squared correlation coefficients for the 100-year test are given for each regression. The black dashed lines indicate 1:1, which would imply a perfect prediction.

response [$\hat{R}(t)$] to the simulated truth, we see that all methods capture internal variability well, explaining between 50% and 80% of the variance in the test.

We do not expect surface temperature to explain all the internal variability in $R$, because not all of the variability in temperature at yearly timescales will translate into a radiative signature. This implies a theoretical maximum performance from using temperature as a predictor. Three distinct approaches (the three regularized regressions, partial least squares, and the CNN) share the maximum value of 80%, which points to a predictability limit for MPI-ESM1.2. Fredericks et al. (2025) find evidence that the 80% limit applies across four climate models and does not seem to improve with additional training data.

Various factors account for the spread in performance among the methods. For OLS, predictive skill is relatively high, but likely falls short of other methods due to overfitting to the training data. The slightly lower skill demonstrated in PC regression suggests the dimensionality reduction in this method does not overcome the overfitting problem, and may even lose some relevant information. MCA can be interpreted as PLS limited to a single covariance component. MCA performs worse than PLS, which uses cross-validation to determine the number of retained components. The remaining methods are less tailored to simply predicting radiative anomalies in internal variability. The Green's function results from equilibrated responses to sustained SST forcing, so it reasonably may not encode transient behavior in internal variability. FDR is designed to incorporate feedbacks across timescales and approximates coupled dynamics when used in the coupled configuration. The atmosphere-only configuration is designed to approximate a Green's function, so might also miss details of internal variability. AMIP ridge regression is trained from SSTs with forced temperature change, which may detract from its ability to predict the unforced variability in the piControl test. However, the uniform warming signal is removed, so a more likely explanation for its lower performance is its training sample of only 145 years, compared to the ~1,000 years of piControl available to the other methods.

In Fig. 1, we show the sensitivity map associated with each of these methods. A sensitivity map describes the global radiative anomaly response to an anomaly in surface temperature at each location. In a linear method, this corresponds to the spatial map of $\beta_i$ coefficients in Eq. 4. The coupled configuration of FDR, the CNN, and AMIP ridge regression have slightly different sensitivity map interpretations. The FDR-coupled sensitivity map represents an operator linking responses to perturbations across all variables and time scales. When testing with a forced experiment, the method aims to estimate the change in TOA radiation at time $t$, given the cumulative change in surface temperature at all previous times $t - \tau$. However, the atmospheric FDR method can be interpreted linearly because it constrains the timescale response to a 1-month time step, essentially only allowing relatively fast atmospheric processes to be captured. In the CNN, the network introduces nonlinearity between layers, so the "gradient" of radiation sensitivity to temperature at each location is dependent on the full SST field; the sensitivity map is unique for each unique map of temperature anomalies. The sensitivity map shown for the CNN in Fig. 1 is the average across all temperature patterns in the out-of-sample internal variability test.

The sensitivity map for AMIP ridge regression predicts only the "pattern component" of the radiative response. The approach first compares the global radiative response in an AMIP simulation to an AMIP-p4K simulation, i.e. an identical simulation with prescribed SSTs, but with a uniform 4 K added to SST at every location. This provides an estimate for the global component of the radiative response to a given global SST change, i.e. $\lambda$. Regression coefficients are determined from an AMIP-piForcing experiment which forces the climate model with observed SSTs from 1870-2014 with forcing fixed to pre-industrial conditions. At each time step, the global mean change in SST is removed from the time-varying spatial SST anomaly, leaving an SST pattern component independent of the global mean warming. The corresponding mean warming component of the radiative response is estimated using the global $\lambda$ derived from the AMIP and AMIP-p4K experiments. The mean warming component is removed from the full radiative time series, leaving a pattern component of radiative response. Ridge regression coefficients are found using only the pattern components of SST and global radiation, so the sensitivity map explains only the pattern component of $R$.

Despite the similarity across their predictions in Fig. 2, different methods are associated with very different patterns of spatial feedbacks. Notably, it is not immediately obvious from how "realistic" a given map looks whether it will have high predictive skill. For example, the ordinary least squares fit can explain 76% of the variance in the out-of-sample test, but has a predominantly noisy sensitivity map. Conversely, MCA and the Green's function have the most contiguously defined features which also overlap with physically-significant regions, but these have the lowest predictive skill. One concern this may raise is whether the specific metric of predictive skill comes from relatively few locations that are difficult to identify, while physical-looking features only emerge from the correlation structure in the temperature field. The sparsity introduced with LASSO and the dimensionality reduction performed in the FDR method attempt to account for this possibility. We pose that predictive skill is not a sufficient measure for whether a method's representation of the pattern effect emerges for the right reasons.

## 5. Application to a global warming scenario

We have described the internal variability comparison in Fig. 2 as an in-distribution test because most methods have been trained on ~1,000 years of internal variability from the same piControl experiment as the test, and the AMIP ridge regression map is also trained on simulated internal variability. We compare these methods to the Green's function and to one another to show their relative ability to capture radiative feedbacks due to internal variability and the different sensitivity maps that lead to their differing in-distribution predictive skill.

One of the most relevant implications of the pattern effect is the discrepancy between the historical pattern of warming and that simulated by climate models, which would lead to a changing radiative feedback in time (e.g., Andrews et al. 2022; Ceppi and Gregory 2017; Armour 2017; Zhou et al. 2016). The sensitivity map of a Green's function allows one to predict what the radiative response to these patterns may be. It has become a common evaluation for the Green's function to test its predictions to patterns of warming in simulated forced warming experiments against the model-simulated truth (Zhou et al. 2017; Dong et al. 2019; Alessi and Rugenstein 2023; Zhang et al. 2023b; Bloch-Johnson et al. 2024). For the same reason, and to compare with the Green's function, several of the statistical methods in this paper have undergone similar tests (Rugenstein et al. 2025; Falasca et al. 2025; Fredericks et al. 2025). Whether sensitivity maps trained on internal variability can capture the feedbacks relevant to predict the forced response is an open question, and for some methods requires more information than is provided by our 1,000-year piControl training protocol. However, for completeness and because it is the natural extension of the problem, we also perform a forcing scenario test with all methods. Green's functions are calculated from equilibrated patch perturbations, while the statistical methods have only seen interannual surface temperature variations, but no forcing. Prior work has suggested that temperature patterns of internal variability are linked to patterns in the forced response (e.g., Dessler 2013; He et al. 2021; Colman and Hanson 2017; Zhou et al. 2015; Davis et al. 2024; Hanke and Proistosescu 2024), so there is reason to believe the statistical methods may be able to identify a forced signal. However, the existing literature is inconclusive about the nature of this link.

We use a step forcing 4x$CO_2$ simulation in MPI-ESM1.2 to compare how each method predicts the forced response (Fig. 1, column two). This is an out-of-distribution test because the training process includes no radiative forcing. Both the Green's function and AMIP ridge regression utilize SST perturbations larger than variability alone, but neither includes a change in $F$.

The simpler linear methods consistently recreate the shape of the forced response but underestimate the magnitude. The coupled version of FDR most closely matches the magnitude of the response, though the transient response is underestimated while the equilibrium response is slightly overestimated. Atmospheric FDR overestimates the response. The CNN strongly underestimates the radiative response, which differs from the results of Rugenstein et al. (2025), who showed that this procedure could estimate the forced response well. However, our training protocol is much more constrained in this comparison where we prioritize a standardized protocol to directly compare methods. Each method may have more optimal performance with different training data (e.g., different amount of training data or temporal frequency) which is out of the scope of the present work. Training data size is critical to machine learning performance (e.g., Hestness et al. 2017; Bahri et al. 2024), and the CNN in Rugenstein et al. (2025) had an order of magnitude more training data than used here. While 1,000 samples are sufficient to predict in-distribution internal variability, the CNN's nonlinearity makes out-of-distribution behavior harder to anticipate. For example, the land mask used in Rugenstein et al. (2025) appears to improve out-of-distribution prediction, but even their CNN underestimates the radiative response by the end of the century.

The Green's function overpredicts the radiative response, which also differs from prior work showing a better agreement (Alessi and Rugenstein 2023, Supporting Information). This is because the version of the MPI-ESM1.2 Green's function here is the raw result from applying the Green's function protocol, whereas the Green's function that better matches the forced radiative response has been calibrated slightly. Following a similar process in Dong et al. (2019), very small negative values in the full Green's function matrix were set to zero, which reduced the negative bias and brought the prediction closer to the model-simulated result (Alessi and Rugenstein 2023, Supporting Information). An alternative method for removing values considered noise from a Green's function would be to apply a significance test to values in the response-function matrix (e.g., Zhang et al. 2023b). When applied to the MPI-ESM1.2 Green's function, this results in an underestimated radiative response (Alessi and Rugenstein 2023, Supporting Information). The AMIP ridge regression method also overestimates the radiative response, which can be explained by its prediction procedure: the sensitivity map in Fig. 1 for AMIP ridge regression only predicts the pattern component of the radiative response, which is assessed from the surface temperature pattern with uniform warming removed. The full prediction includes both this pattern component and the radiative component from uniform warming, which uses a global $\lambda$ estimated from the difference between AMIP and AMIP-p4K. In this case, $\lambda$ is approximately -1.8 W m$^{-2}$ K$^{-1}$, which overestimates the response. This is likely because the positive ice-albedo feedback is not included in amip-p4k, where sea ice is fixed at observed values.

We could also estimate the global feedback parameter, $\lambda$, by taking the regression of global radiation onto global temperature in the piControl training data. This is a null hypothesis relative to methods designed to account for the pattern effect, because this approach uses no spatial information. The value of this feedback parameter is -1.34 W m$^{-2}$ K$^{-1}$. The prediction from the null hypothesis matches the simulated magnitude of the radiative response more closely than that of any of the pattern effect methods (not shown). However, this result is strongly model dependent. Bloch-Johnson et al. (2020) have shown that the internal variability global feedback parameter in the MPI-ESM1.2 climate model does well at predicting the forced response, but this is not true of all models.

The forced, long-term climate response behaves differently than responses to interannual temperature fluctuations. Relating internal variability to the forced response is a well-established effort in climate dynamics (e.g., Leith 1975; Chung et al. 2010; Dessler 2010, 2013; Dessler and Forster 2018; Brient and Schneider 2016; Lutsko and Takahashi 2018; Lutsko et al. 2021; He et al. 2021; Schlund et al. 2020; Sherwood et al. 2014; Uribe et al. 2022; Zhou et al. 2015; Davis et al. 2024), and the range of 4x$CO_2$ predictions in Fig. 1 emphasizes the challenge in relating the two. Additionally, most methods here assume linearity, but it has been shown that the heterogeneity of temperature or heat uptake patterns has a secondary effect to the pattern effect (Rugenstein et al. 2016; Bloch-Johnson et al. 2024). This is also related to the effect of patch size, discussed in the next section. More evidence for nonlinearity comes from patch forcing experiments that show the superposition of two independent patch responses does not equal the response when both patches are perturbed at the same time (Williams et al. 2023; Quan et al. 2024). Nonlinearities enhance the difference between the training data and the out-of-distribution 4x$CO_2$ test, contributing to the discrepancies among the methods' predictions. Even the CNN, which can account for nonlinearities, tends to become linear out-of-distribution.

We highlight that our goal here is not to make numerical predictions or argue for one or the other method to be superior at out-of-distribution predictions, but to assemble and discuss all methods. The methods diverge despite identical training and testing data, so there is clearly methodological uncertainty in the various predictions, in part due to the different assumptions inherent to each method. The comparison provided here does not reveal whether the methods that appear “better” in Fig. 1 are in fact better at identifying the true underlying physics, or whether they are better due to sampling variability and/or methodological “luck”. For example, a given method may perform well by coincidence (as is true for global $\lambda$ as discussed above) even in the extreme case where the forced response is not learnable from internal variability. We leave sensitivity testing to future work to determine whether the divergence primarily stems from the limits of each method, the limits of training on internal variability, or the applicability of internal variability to the forced response.

## 6. Converting statistical methods to the Green’s function protocol

Most methods agree on the basic structure found across Green’s function experiments, with negative feedbacks in the west Pacific and Brazilian coast, and positive feedbacks in the subtropical highs of the east Pacific and Namibian coast. This is in line with the previously discussed mechanism whereby warming in subsidence regions reduces low clouds, but warming in convective regions increases low cloud cover through remote effects on tropospheric stability. The major exception to these features is the coupled version of FDR. This is the only method that allows the surface temperature to change in response to the applied temperature anomalies. While the other methods can only capture atmospheric processes, this allows the ocean to respond as well over slower timescales. The ocean response likely explains the diverging sensitivities, though further work is necessary to understand by what processes.

Despite their broad agreement with the Green’s function, sensitivity maps produced by the statistical methods frequently include spatial features not present in the Green’s function map. One prominent difference is the treatment of land regions. Prescribing land surface temperature in climate models has been historically difficult (e.g., Andrews et al. 2021; Bloch-Johnson et al. 2024). Some experiments have been performed with perturbed land surface temperatures (Ackerley and Dommenget 2016; Ackerley et al. 2018; Zhang et al. 2026), so Green’s functions may include land in the future, but they have thus far perturbed only the sea surface temperature. Similarly, the uniform +4 K warming required for AMIP ridge regression has only once to our knowledge been implemented in an atmospheric global climate model with prescribed land surface temperatures (Ackerley et al. 2018). In contrast, we train and test against near-surface air temperature ("tas") in this comparison. All of the statistical methods are able to use values both over land and ocean as predictors, which show up as contributions that the numerical methods cannot capture. When Thompson et al. (2025) applied MCA directly to the last 20 years of observations, they found that temperature variability over land areas accounts for ~70% of the global negative feedback. Fredericks et al. (2025) found similar feedback patterns in the observations with ridge regression and elastic net.

Where methods directly disagree with the Green's function, it introduces the question of which better represents the underlying physical feedbacks. For example, the most prominent area of disagreement with the Green's function in several methods is an equatorial band of positive feedback extending from the east to the central Pacific. This feature is reminiscent of the ENSO temperature anomaly, so a possible explanation might be that the patch forcings in the Green's function protocol never include a spatially coherent pattern of ENSO variability. In contrast, the statistical methods are trained on internal variability patterns including ENSO. However, another possibility for this difference is the spatial heterogeneity allowed by the different methods. The Green's function protocol calls for patches that span tens of degrees (Bloch-Johnson et al. 2024), whereas the methods showing a positive equatorial feedback band calculate feedbacks on the gridpoint scale.

To test whether statistical methods are translatable, we use the trained convolutional neural network rather than a numerical model to perform a Green's function experiment. We apply the CNN to the SST boundary conditions of the Green's function protocol. That is, we use the CNN to predict $R$ from the control SST and for each patch perturbation separately (e.g., Fig 3b). As in the Green's function, we then use the difference between the patch simulations and the control to generate a linear sensitivity map. This results in a CNN-generated Green's function sensitivity map (Fig. 3c). We use the CNN because it performs well at the internal variability test and encodes slightly more information than the regression maps through its nonlinearity, making it a good representative for methods trained on internal variability. The CNN-generated Green's function sensitivity map (Fig. 3c) looks more similar to the GCM-simulated Green's function (reproduced in Fig. 3d) than to the sensitivity map as assessed from internal variability (reproduced in Fig. 3a). The CNN is reproducing the behavior in the climate model that leads to the simulated Green's function sensitivity map. The equatorial central Pacific positive feedback, for example, is absent in Fig. 3c, as is most of the smaller-scale heterogeneity in Fig. 3a.

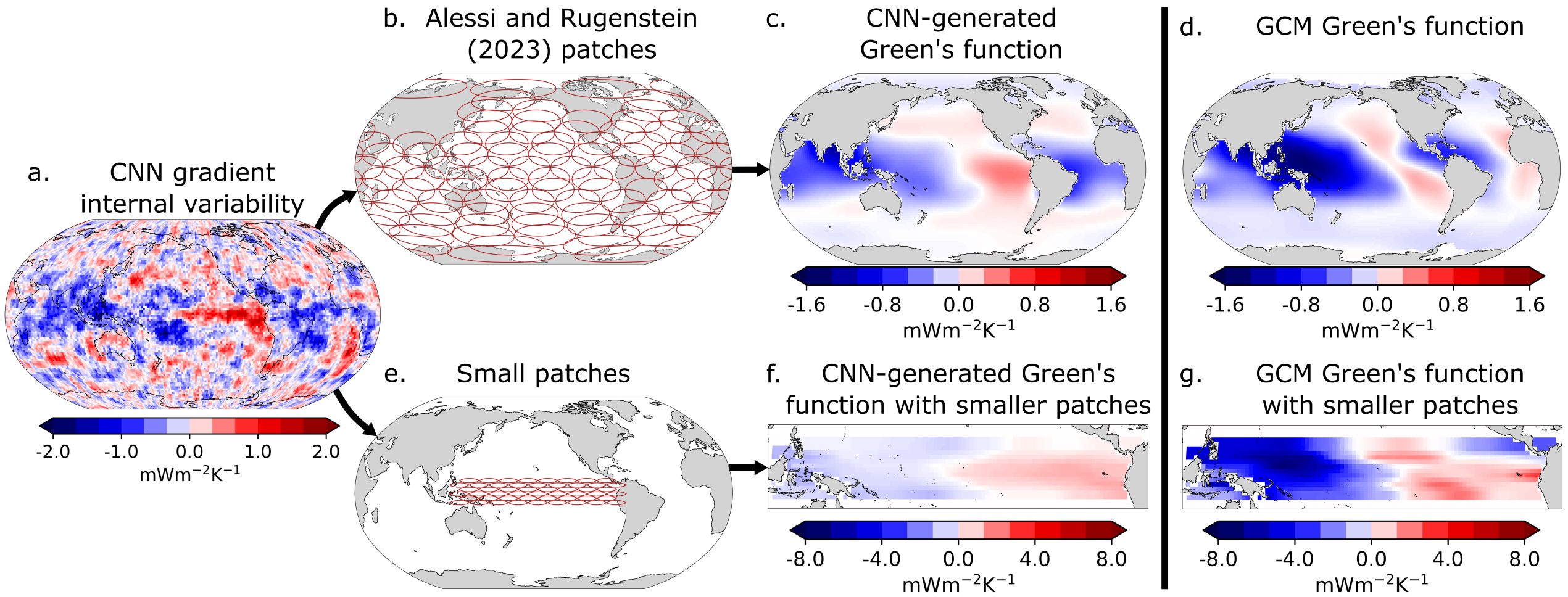

Fig. 3. Comparison of Green's function experiments performed with the patch-size protocol used in Alessi and Rugenstein (2023) (top row) vs. with a smaller patch size in the equatorial Pacific (bottom row). a) Average gradient for the convolutional neural net trained on internal variability, identical to the CNN sensitivity map in Fig. 1. b) Patches outlined at the half-patch width for the global protocol, c) sensitivity map produced by using the trained CNN to predict radiative responses to patch perturbations following the Green's function protocol, performed on the patches in (b), and d) sensitivity map produced from patch perturbations in MPI-ESM1.2. The map in (d) is identical to the Green's function sensitivity map in Fig. 1. e) Smaller patches outlined at the half-patch width for a new experiment limited to the equatorial Pacific. f,g) As in (c,d), but using the patch perturbations outlined in (e). Note the difference in scale resulting from patch size.

To test whether the smaller spatial scales and larger local magnitudes found by many statistical methods bear truth, we compute an additional GCM Green's function experiment with smaller patches (Fig 3g). Due to computational costs, we perform them only in the equatorial Pacific (Fig 3e). From the GCM simulations, we find smaller scale features when using the smaller patch perturbations than are present for the standard patch experiments. We also perform an analogous experiment on the smaller patches with the trained CNN, as we did with standard patches (Fig.

3f). As with the GCM, the CNN-generated Green's functions reveal smaller-scale spatial structure for small patches than for standard patches.

The responses are not identical, but both have positive feedbacks reaching farther into the central Pacific and more heterogeneity in the large West Pacific negative feedback region. Though still more homogeneous than feedbacks assessed at the grid cell level (e.g., Fig. 3a), these experiments show that the typical Green's function protocol obscures some of the smaller spatial scales that would be present from smaller patch sizes. The smooth, large-scale features in the Green's function sensitivity map are not a property of the atmospheric climate model. This does not necessarily mean smaller patches improve the Green's function. Bloch-Johnson et al. (2024) observe that small patches appear to worsen the nonlinearity in Green's function experiments by increasing the disparity between patch-warming and patch-cooling experiments. They also find that smaller patches lead to stronger responses. Though both the CNN and GCM Green's functions exhibit this behavior (Figs. 3f and 3g), it is unclear why the CNN-generated map has a weaker response. There may be relatively few examples in the internal variability training data with large, highly-localized anomalies. This would mean the CNN does not activate the same nonlinearities observed by Bloch-Johnson et al. (2024). Through these comparisons, statistical methods help to optimize the Green's function protocol.

## 7. Summary and looking forward

The methods all highlight different elements of the same problem. For example, the temperature field as a whole covaries with global radiation (MCA), not all local temperatures affect global radiation in a strong way (LASSO, elastic net), the system evolves around dominant patterns of variability (PC regression), both the ocean and atmosphere respond to warming, leading to different feedbacks at different timescales (FDR), temperature patterns can affect $R$ nonlinearly (CNN), and directly causal responses to artificial perturbations (Green's function) may differ from correlations in the naturally-emerging temperature field (AMIP ridge regression). No single method incorporates these disparate features yet, but all give insights into properties of the system connecting surface temperature and radiation.

Most of the methods implement some variation of multilinear regression (OLS, MCA, LASSO, elastic net, ridge, PC regression, PLS). These have the common advantage of inferring sensitivity maps from coupled systems with relatively common, simple to implement procedures. Some, like LASSO and elastic net, also introduce sparsity to the solution. Sparsity can highlight regions most relevant to prediction, but its incompleteness may also be a disadvantage for some applications. The common disadvantage to these methods is that they only diagnose instantaneous sensitivities from correlations within the data, leaving causality ambiguous. The relative disadvantages among the multilinear regression variants relate to how they handle degrees of freedom in the data and their tendencies towards overfitting. In several cases, this requires fitting parameters through cross-validation. AMIP ridge regression is also based in multilinear regression, but the setup incorporates observed SST patterns and historical forcing, potentially making it more relevant to predicting a forced response. The disadvantage of this method is that it requires specific forcing simulations, so cannot be assessed from internal variability like the other regression methods.

The FDR method has an additional advantage of incorporating lagged responses to capture feedbacks across different timescales. This also allows it to behave more like a coupled system than the other methods. FDR's complexity has the disadvantage that it is harder to interpret in the coupled version, though the atmospheric version approximates Green's function behavior. Also difficult to visualize are the sensitivity maps produced by the CNN, which are themselves dependent on the pattern in surface temperature. Our results suggest the neural network needs more data than the other methods to predict a realistic forced response, and training requires the selection of several hyperparameters. However, the CNN has the advantage of accounting for nonlinearities in the temperature field, and links this application to the rapidly improving machine learning field.

The Green's function method solely has the advantage of being causal, in that every assessed atmospheric response necessarily emerges from surface temperature perturbations. However, by prescribing surface temperature, the Green's function protocol cannot capture coupled behavior. Distinguishing partial contributions also requires a large number of dedicated simulations, while existing coupled simulations can be used by data-driven methods. The quantified values in the sensitivity map for the Green's function also have a slightly different interpretation. They show the direct contribution, through the particular patch protocol employed, that perturbing temperature will have on $R$. With the data-driven methods, the values are essentially regression coefficients that collectively predict $R$. The assessed feedback values vary widely in magnitude (Fig. 1) due to both horizontal resolution and sparsity. When it comes to quantitatively predicting $R$, we find that the Green's function does not perform as well as statistical methods for internal variability, but does reasonably well in prior investigations at predicting the forced response after being calibrated/tuned. Our results do not conclusively determine how well statistical methods can predict internal variability, but we have found evidence for a maximum $R^2$ of approximately 80% between prediction and truth. The spread in predictions for the 4x$CO_2$ scenario demonstrates that quantitatively predicting the forced response is less constrained. Future investigations should also consider if $R^2$ between actual and predicted internal variability and the relative overall magnitude of the forced response are sufficient criteria for whether a particular method is "good" at predicting $R$.

Among the advantages from being applied to coupled systems, statistical methods open the door for estimating sensitivity maps from observations. The numerical Green's function approach has no observational equivalent, but most of these new methods can be applied to observations in some way. AMIP ridge regression trains on observed sea surface temperatures in a blend of numerical and observational data. The limited observational data proves a challenge for FDR and the CNN, which structurally require larger training sets. However, ridge regression, LASSO, elastic net, and MCA have already been applied directly to observations (Thompson et al. 2025; Fredericks et al. 2025). Fredericks et al. (2025) also found evidence that the existing 24 years of reliable satellite measurements for $R$ do not represent the full range of variability reachable in the system. Uribe et al. (2024) show that using internal variability for an observational constraint on the forced response will require around 50 years of continuous satellite data. The exact number of years needed to directly generate a sensitivity map will depend in part on the method used. For example, a convolutional neural network requires orders of magnitude more training data than linear methods, rendering CNNs less useful in observational analyses of the pattern effect. It may be impractical to use some of these methods with observations. Their usefulness depends on how uncertainty is quantified, and what level of uncertainty is considered appropriate to have represented the system. Observations also introduce the question of how to remove the forcing, which is uncertain in the real world. Even if we trust that a particular method represents the observations accurately, our comparison across the 4x$CO_2$ test suggests that predictive skill in internal variability may not translate to predictive skill for the forced response.

One way to circumvent observational data limitations would be to rely more heavily on observationally-derived products. ECMWF ERA5 reanalysis (Hersbach et al. 2020) estimates net radiative flux back to 1940, though extending a record with ERA5 alone would introduce significant measurement uncertainty (e.g., Loeb et al. 2022; Uribe et al. 2024). Alternatively, recent developments in machine learning models have led to models that can emulate ERA5, such as the Ai2 Climate Emulator (ACE; Watt-Meyer et al. 2023, 2025). ACE-ERA5 can be run with prescribed sea surface temperatures, which means patch perturbations are possible. Van Loon et al. (2025b) used ACE-ERA5 with the Green's function protocol and found that it qualitatively matched Green's functions in global climate models. Because the behavior has been trained on ERA5, this is to some degree an observationally-derived Green's function.

Beyond the pattern effect, our comparison would also be useful to anyone identifying the patterns in one variable related to the response in another. We have shown how methods developed primarily to identify spatial patterns (maximum covariance analysis, the Green's function, and the fluctuation-dissipation relation) can be related directly to methods developed primarily to solve the multilinear regression problem (ordinary least squares, regularized regression, principal component regression, partial least squares, and the convolutional neural network). Any problem suited to a single one of these methods can therefore benefit from the sensitivity maps or predictive abilities of the others. An extension of these methods would be to explore the spatial structure in $R$, in which the patterns in temperature would relate to local radiative time series. Because this requires training on each local $R$ separately, which significantly increases training time in some methods like the CNN, we leave this to future work.

What we show here is a case study exploring wide-ranging interpretations of the pattern effect. They expand our ability to predict climate change through our understanding of $R$ in the following ways:

1. Data-based methods offer several new ways to explore the pattern effect relative to Green's function numerical simulations. These include the abilities to learn from coupled atmosphere-ocean simulations, to use land surface temperatures, and to apply directly to observations. Here we have provided both a qualitative comparison of sensitivity maps and a quantitative comparison of their ability to predict $R$ in test cases across these methods. We have summarized the advantages and disadvantages of the methods in Fig. 1. We emphasize that each method has considerable complexity that is beyond the scope of the comparison provided here. For further details, the reader is referred to further discussion of each method in the relevant source literature.

2. These methods highlight regional influences that underlie physical processes. Broadly, the methods support the existing evidence for negative feedbacks in regions of deep convection and positive feedbacks in regions of subsidence and climatological stratocumulus. More narrowly, some hint at a larger role for ENSO, an inverted response when the ocean is included, or the insignificance of large parts of the globe. All implications can and should be explored in process-oriented studies.

3. The methods allow direct prediction of the forced response (Fig. 1). The methods predict a wide range of radiative responses for the specific task in this paper, predicting radiative response in one model's 4x$CO_2$ simulation. However, they can be improved and calibrated when not constrained by the design of our test. Predictions in this work should be seen as a starting point rather than a limitation.

*Acknowledgments.* LF was supported by the NSF Climate and Large-Scale Dynamics program under AGS-2116186 and AGS-2233673, and by a NASA FINESST grant 80NSSC24K0024. MR was supported by the National Aeronautics and Space Administration under Grant No. 80NSSC21K1042, NSF RISE-2530919, and NSF 014298-0002. D. W. J. T. was supported by the National Aeronautics and Space Administration (NASA) under 80NSSC23K0113; the National Science Foundation (NSF) CLD Program under AGS-2116186; and NSFGEO-NERC 2507587. SVL was supported by NSF RISE-2530919. FF was supported by Schmidt Sciences, LLC, through the M$^2$LInES project. RBF was supported by the Eric and Wendy Schmidt AI in Science Postdoctoral Fellowship, a program of Schmidt Sciences. PC was supported by UK Research and Innovation (UKRI) under the UK government's Horizon Europe funding Guarantee (grant EP/Y036123/1). PC was additionally supported through UK Natural Environmental Research Council (NERC) grants NE/V012045/1 and NE/T006250/1. QW received support from the European Research Council under the European Union's Horizon 2020 research and innovation programme (grant agreement No. 786427, project "Couplet").

*Data availability statement.* Preindustrial control and 4x$CO_2$ data for MPI-ESM 1.2 are available through the LongRunMIP project, https://www.longrunmip.org/. The MPI-ESM Green's function is available from Alessi and Rugenstein (2023). The fluctuation-dissipation relation was implemented as in Falasca et al. (2025). AMIP ridge regression was implemented as in Kang et al. (2023). The convolutional neural network was implemented as in Rugenstein et al. (2025). Ordinary least squares was implemented as in Bloch-Johnson et al. (2020). Maximum covariance analysis and the three regularized regression methods were implemented as in Fredericks et al. (2025). Principal component regression as applied to this paper is available at https://doi.org/10.5281/zenodo.18745163 (Wu 2026). Partial least squares as applied to this paper, the Green's function with small patches, data for test results and sensitivity maps, and plotting scripts are available at https://doi.org/10.5281/zenodo.18840026 (Fredericks 2026).